\documentclass[12pt]{article}
\usepackage{latexsym}
\usepackage{exscale}
\usepackage{psfig}
\newcommand{\mbb}[1]{\mbox{\boldmath $#1$}}

\begin{document}

\begin{center}
\LARGE
Entanglement degradation of a two-mode squeezed vacuum in absorbing
and amplifying optical fibers\\[2ex]
\large
Stefan~Scheel$^1$, Toma\v{s}~Opatrn\'{y}$^{1,2}$, and
Dirk-Gunnar~Welsch$^1$\\[1ex]
\normalsize 
$^1$Theoretisch-Physikalisches Institut, 
Friedrich-Schiller-Universit\"{a}t Jena,\\
Max-Wien-Platz 1, D-07743 Jena, Germany\\
$^2$Department of Theoretical Physics, Palack\'{y} University,\\ 
Svobody 26, 771~46 Olomouc, Czech Republic\\[4ex] 
\end{center}
\thispagestyle{empty}

\begin{abstract}
Applying the recently developed formalism of quantum-state 
transformation at absorbing dielectric four-port devices
[L.~Kn\"oll, S.~Scheel, E.~Schmidt, D.-G.~Welsch, and A.V.~Chizhov, 
Phys. Rev. A {\bf 59}, 4716 (1999)], we calculate the quantum state 
of the outgoing modes of a two-mode squeezed vacuum transmitted 
through optical fibers of given extinction coefficients. 
Using the Peres--Horodecki separability criterion for
continuous variable systems [R.~Simon, Phys. Rev. Lett. {\bf 84}, 
2726 (2000)], we compute the maximal length of transmission 
of a two-mode squeezed vacuum through an absorbing system 
for which the transmitted state is still inseparable. Further,
we calculate the maximal gain for which inseparability can be 
observed in an amplifying setup. Finally, 
we estimate an upper bound of the entanglement preserved 
after transmission through an absorbing system. The results show that
the characteristic length of entanglement degradation drastically 
decreases with increasing strength of squeezing.
\end{abstract}

\section{Introduction}

Recently there has been increasing interest in the use of
entangled continuous variable systems in quantum communication
\cite{Braunstein98}. The most prominent
example is the two-mode squeezed vacuum which in the
Fock basis reads as
\begin{equation}
\label{0.1}
|\psi\rangle 
= {\rm e}^{\zeta \left( \hat{a}_1^\dagger
\hat{a}_2^\dagger - \hat{a}_1 \hat{a}_2 \right)} |00\rangle
= \sqrt{1-q^2} \sum\limits_{n=0}^\infty
q^n \,|nn\rangle 
\end{equation}
[$q$ $\!=$ $\!\tanh\zeta$, $\zeta$ real] and whose Wigner function is
a Gaussian. When transmitted through a noisy channel,
such as optical fibers of given extinction coefficients,
the resulting state will in general
be mixed. For further use of the state in some quantum
communication experiment the question of the entanglement that 
is available after the transmission arises.
Real entanglement measures, however, are difficult to compute and are
mostly known only numerically. A typical example is the entanglement 
measure based on the relative entropy where the distance 
of the density matrix of the state under consideration to the set of all 
separable density matrices must be computed \cite{Vedral9798}. 
In practice, this becomes impossible for states like a 
two-mode squeezed vacuum with reasonable strength of squeezing,
because for comparable numerical accuracy the Hilbert space can only be
truncated at higher dimension with increasing squeezing strength. 

Here, we proceed in a different direction. 
Starting from the quantum-optical input-output relations 
of light at absorbing dielectric four-port devices \cite{Gruner96b},
we first present the basic formulas for determining the
output quantum state from the input quantum state and the 
characteristic transmission and absorption matrices of the
devices (Sec.~\ref{state}). We then apply the input-output 
formalism to the calculation of the output Wigner function 
observed when two modes that are initially 
prepared in a squeezed vacuum are transmitted through absorbing fibers. 
Using the Peres-Horodecki separability criterion for Gaussian states 
\cite{Simon00}, we calculate the maximal length of transmission 
for which the output state is still inseparable in principle
(Sec.~\ref{sep}). Calculating the density matrix of the 
transmitted light in the Fock basis, on applying the formalism 
developed in \cite{Knoll99}, we finally extract some pure 
state from the output density matrix, use the convexity property of 
the entanglement measure \cite{Scheel00}, and derive an estimate of 
the amount of entanglement available after transmission
(Sec.~\ref{estimate}).

\section{Quantum-state transformation}
\label{state}

Let us first consider the quantum-optical input-output relations and
the corresponding quantum-state transformation formulas for light
at dispersing and absorbing dielectric four-port devices.
\begin{figure}[h]
\hspace*{5cm}
\psfig{file=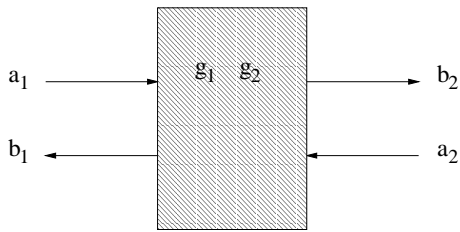,width=6cm}
\caption{\label{slab} 
\protect\parbox[t]{.67\textwidth}{Scheme of the dielectric 
four-port device, with \protect$\hat{a}_i$ $\!=$ $\!\hat{a}_i(\omega)$ 
[\protect$\hat{b}_i$ $\!=$ $\!\hat{b}_i(\omega)$] and
\protect$\hat{g}_i$ $\!=$ $\!\hat{g}_i(\omega)$ being the 
destruction operators of the input [output] photons and the
device excitations respectively.}
}
\end{figure}
We restrict ourselves to a quasi one-dimensional scheme, as depicted in 
Fig.~\ref{slab}, in which the dielectric device is surrounded by vacuum. 
Applying the formalism developed in \cite{quant}, we 
quantize the electromagnetic field in the presence of the device  
by means of a Green function representation of the field 
and introduction of bosonic fields playing 
the role of the collective excitations of the field, the dielectric 
matter, and the reservoir. It turns out that outside
the device the usual mode expansion applies, with the $\hat{a}_i(\omega)$
and $\hat{b}_i(\omega)$ in Fig.~\ref{slab} being respectively the photonic 
operators of the incoming and outgoing plane waves at frequency $\omega$.
The $\hat{g}_i(\omega)$ in the figure are the bosonic operators of the 
device excitations and play the role of noise forces associated which 
absorption. It then follows that the action of the dielectric device on 
the incoming radiation can be described by quantum optical input-output 
relations which, in fact, are nothing but a suitable 
rewriting of the corresponding one-dimensional Green function. 
Let us introduce the compact two-vector notation 
$\hat{\bf a}(\omega)$, $\hat{\bf b}(\omega)$, and $\hat{\bf
g}(\omega)$ for the field and device operators respectively. The
input-output relations can then be written in the compact form
\cite{Gruner96b} 
\begin{equation}
\label{1.1}
\hat{\bf b}(\omega) = {\bf T}(\omega) \hat{\bf a}(\omega)
+{\bf A}(\omega) \hat{\bf g}(\omega),
\end{equation}
where the characteristic transmission and absorption matrices 
${\bf T}(\omega)$ and ${\bf A}(\omega)$, respectively, satisfy the 
energy-conservation relation
\begin{equation}
\label{1.2}
{\bf T}(\omega) {\bf T}^+(\omega) +{\bf A}(\omega) {\bf A}^+(\omega)
= {\bf I}. 
\end{equation}
Equations~(\ref{1.1}) and (\ref{1.2}) are
valid for any frequency. They allows us to construct the output 
operators from the input operators over the whole frequency range.

The second step in the quantum-state transformation corresponds
to an open-systems approach. Suppose the incoming field is prepared in
some state of the Hilbert space ${\cal H}_{\rm field}$ and the device
(including the reservoir) is initially prepared in some state of the 
Hilbert space ${\cal H}_{\rm device}$. In the
full Hilbert space, which is the tensor 
product ${\cal H}_{\rm field} \otimes {\cal H}_{\rm device}$,
a unitary operator transformation can then be
constructed, whereas in the space ${\cal H}_{\rm field}$ it could not
due to the dissipation processes.
Let us define the four-vector operators
\begin{equation}
\label{1.3}
\hat{\mbb{\alpha}}(\omega) =
{\hat{\bf a}(\omega) \choose \hat{\bf g}(\omega)} \,, \quad
\hat{\mbb{\beta}}(\omega) =
{\hat{\bf b}(\omega) \choose \hat{\bf h}(\omega)},
\end{equation}
where $\hat{\bf h}(\omega)$ is some auxiliary (two-vector) 
bosonic device operator. Then, the input-output relation (\ref{1.1}) can 
be extended to a four-dimensional transformation
\begin{equation}
\label{1.4}
\hat{\mbb{\beta}}(\omega) = \mbb{\Lambda}(\omega)
\hat{\mbb{\alpha}}(\omega) \,,
\end{equation}
with $\mbb{\Lambda}(\omega)$ $\in$ SU(4) \cite{Knoll99}. Explicitly,
\begin{equation}
\label{1.4.2}
\mbb{\Lambda}(\omega) =
\left( \begin{array}{cc}
{\bf T}(\omega) & {\bf A}(\omega) \\ -{\bf S}(\omega) {\bf
C}^{-1}(\omega) {\bf T}(\omega) & {\bf C}(\omega) {\bf S}^{-1}(\omega) 
{\bf A}(\omega)
\end{array} \right)
\end{equation}
with the commuting positive Hermitian matrices
\begin{equation}
\label{1.4.3}
{\bf C}(\omega) = \sqrt{{\bf T}(\omega){\bf T}^+(\omega)} \,, \quad
{\bf S}(\omega) = \sqrt{{\bf A}(\omega){\bf A}^+(\omega)} \,.
\end{equation}
Hence, there is a unitary operator transformation
\begin{equation}
\label{1.5}
\hat{\mbb{\beta}}(\omega) = \hat{U}^\dagger \hat{\mbb{\alpha}}(\omega) 
\hat{U}
\end{equation}
where
\begin{equation}
\label{1.6}
\hat{U} = \exp\!\left\{ -i\int\limits \!{\rm d}\omega \left[
\hat{\mbb{\alpha}}^\dagger(\omega) \right]^T \mbb{\Phi}(\omega)
\hat{\mbb{\alpha}}(\omega) \right\}
\end{equation}
and
\begin{equation}
\label{1.7}
\mbb{\Lambda}(\omega) = {\rm e}^{-i\mbb{\Phi}(\omega)} \,.
\end{equation}
Note that $\hat{U}$ acts in the product space ${\cal H}_{\rm field}
\otimes {\cal H}_{\rm device}$. Given a density operator
$\hat{\varrho}_{\rm in}$ of the input
quantum state as a functional of $\hat{\mbb{\alpha}}(\omega)$, the
density operator of the output quantum state is obtained by a unitary
transformation with the operator $\hat{U}$ from Eq.~(\ref{1.6}) and
projecting back onto the Hilbert space ${\cal H}_{\rm field}$. Hence,
\begin{equation}
\label{1.8}
\hat{\varrho}^{({\rm F})}_{\rm out} = {\rm Tr}^{({\rm D})} \left\{
\hat{U} \hat{\varrho}_{\rm in} \hat{U}^\dagger \right\} 
= {\rm Tr}^{({\rm D})} \left\{ \hat{\varrho}_{\rm in} \left[
\mbb{\Lambda}^+(\omega) \hat{\mbb{\alpha}}(\omega),
\mbb{\Lambda}^T(\omega) \hat{\mbb{\alpha}}^\dagger(\omega) \right]
\right\} \,,
\end{equation}
where ${\rm Tr}^{({\rm D})}$ means tracing with respect to the device
variables. Note, that the difference to usually considered
open-systems theories is provided by the fact that we actually know
how the dissipative environment (e.g., a dispersing and absorbing fiber)
acts on our quantum states. 

Let us briefly comment on amplifying devices. In contrast to absorbing
devices, we have now to insert the noise {\it creation} operators
$\hat{g}_i^\dagger(\omega)$ into the input-output relation
(\ref{1.1}) [i.e., $\hat{\bf g}(\omega)$ $\!\to$ 
$\!\hat{\bf g}^\dagger(\omega)$]. The relation (\ref{1.2}) then changes to
\begin{equation}
\label{1.9}
{\bf T}(\omega) {\bf T}^+(\omega) -{\bf A}(\omega) {\bf A}^+(\omega)
= {\bf I} \,,
\end{equation}
where ${\bf A}(\omega)$ plays the role of the gain matrix.
Further, the 4$\times$4-matrix $\mbb{\Lambda}(\omega)$ becomes
an element of the noncompact group SU(2,2).

\section[Separability criterion]{Application of the Peres--Horodecki
separability criterion}
\label{sep}

\begin{figure}[h]
\hspace*{4cm}
\psfig{file=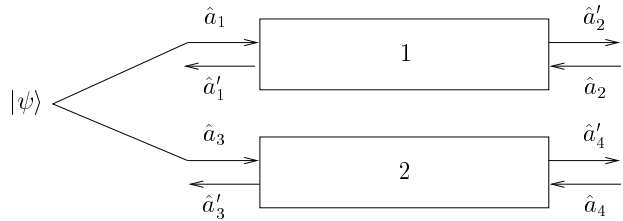,width=8cm}
\caption{\label{fourport}
\protect\parbox[t]{.67\textwidth}{ 
A two-mode input field prepared in the quantum state $|\psi\rangle$
is transmitted through two absorbing (amplifying) dielectric 
four-port devices, $\hat a_1,\hat a_3$ ($\hat a_2',\hat a_4'$) 
being the photonic operators of the relevant input (output) modes.}
}
\end{figure}

Let us consider a two-mode squeezed vacuum 
which is transmitted through a noisy
communication channel such as two absorbing (amplifying) 
dielectric four-port devices as sketched in Fig.~\ref{fourport}, 
the characteristic
transmission and absorption (gain) matrices of the devices being
${\bf T}^{(i)}(\omega)$ and ${\bf A}^{(i)}(\omega)$ respectively
($i$ $\!=$ $\!1,2$).
In Eq.~(\ref{0.1}), the two-mode squeezed vacuum is given
in the Fock basis. Equivalently, it can be expressed in terms of 
the Gaussian Wigner function 
\begin{equation}
\label{3.2}
W(\mbb{\xi}) = \left(4\pi^2 \sqrt{\det \mbb{V}}\right)^{-1}
\exp\!\left\{ -{\textstyle\frac{1}{2}} 
\mbb{\xi}^T \mbb{V}^{-1} \mbb{\xi} \right\},
\end{equation}
where $\mbb{\xi}$ is a four-vector whose elements are the 
quadrature-component variables $q_1,p_1,q_3,p_3$, and 
$\mbb{V}$ is the $4\times 4$ variance matrix of
the Wigner function,
\begin{equation}
\mbb{V}  =  
\left( \begin{array}{cc}
{\bf X} & {\bf Z} \\ {\bf Z}^T & {\bf Y} 
\end{array} 
\right)
=  
\left( 
\begin{array}{cccc}
c/2&0&-s/2&0\\0&c/2&0&s/2\\-s/2&0&c/2&0\\0&s/2&0&c/2
\end{array} 
\right)
\label{3.4}
\end{equation}
($c$ $\!=$ $\!\cosh 2\zeta$, $s$ $\!=$ $\!\sinh 2\zeta$). Transmitting
the two-mode squeezed vacuum through the four-port devices at some
temperatures $\vartheta_i$, the Wigner function of the transformed
state is again a Gaussian. Using the input-output relations
for absorbing (amplifying) devices as given in Sec.~\ref{state},
we can easily transform the input variance matrix 
(\ref{3.4}) to obtain the output variance matrix. The result is
\begin{eqnarray}
\label{3.6}
\langle \hat{a}_2 {}^{\!\!'\dagger} \hat{a}_2' \rangle_s 
&=&
|T_{21}^{(1)}|^2 \langle \hat{a}_1^\dagger \hat{a}_1 \rangle_s
+|T_{22}^{(1)}|^2 \langle \hat{a}_2^\dagger \hat{a}_2 \rangle_s
\nonumber \\ &&\hspace{2ex}
+\,|A_{21}^{(1)}|^2 \langle \hat{g}_1 {}^{\!\!(1)\dagger}
\hat{g}_1^{(1)} \rangle_s 
+|A_{22}^{(1)}|^2 \langle \hat{g}_2 {}^{\!\!(1)\dagger}
\hat{g}_2^{(1)} \rangle_s 
\,, \nonumber\\[.5ex]
\langle \hat{a}_4 {}^{\!\!'\dagger} \hat{a}_4' \rangle_s 
&=&
|T_{21}^{(2)}|^2 \langle \hat{a}_3^\dagger \hat{a}_3 \rangle_s
+|T_{22}^{(2)}|^2 \langle \hat{a}_4^\dagger \hat{a}_4 \rangle_s
\nonumber \\ &&\hspace{2ex}
+\,|A_{21}^{(2)}|^2 \langle \hat{g}_1 {}^{\!\!(2)\dagger}
\hat{g}_1^{(2)} \rangle_s 
+|A_{22}^{(2)}|^2 \langle \hat{g}_2 {}^{\!\!(2)\dagger}
\hat{g}_2^{(2)} \rangle_s 
\,, \nonumber\\[.5ex]
\langle \hat{a}_2' \hat{a}_4' \rangle 
&=&
T_{21}^{(1)} T_{21}^{(2)} \langle \hat{a}_1 \hat{a}_3 \rangle 
\end{eqnarray}
(the subscript $s$ refers to symmetric operator ordering as
required for the Wigner function)
which can also be rewritten as correlations of the quadratures.
Introducing the abbreviating notation $T_{21}^{(i)}$ $\!\equiv$
$\!T_i$ and $T_{22}^{(i)}$ $\!\equiv$ $\!R_i$, we arrive at the 
following elements of the output variance matrix:
\begin{equation}
\label{3.7a}
X_{11}=X_{22}
= {\textstyle\frac{1}{2}}c|T_1|^2 + {\textstyle\frac{1}{2}} |R_1|^2 
+\sigma \left( n_{{\rm th}\,1}+{\textstyle\frac{1}{2}} \right) 
\left( 1-|T_1|^2-|R_1|^2 \right),
\end{equation}
\begin{equation}
\label{3.7b} 
Y_{11}=Y_{22}
= {\textstyle\frac{1}{2}}c|T_2|^2 +{\textstyle\frac{1}{2}} |R_2|^2 
+\sigma \left( n_{{\rm th}\,2}+{\textstyle\frac{1}{2}} \right) 
\left( 1-|T_2|^2-|R_2|^2 \right),
\end{equation}
\begin{equation}
\label{3.7c}
Z_{11} = -Z_{22} 
= -{\textstyle\frac{1}{2}}s {\rm Re}\,(T_1 T_2),
\end{equation}
\begin{equation}
\label{3.7d}
Z_{12} = Z_{21}
=- {\textstyle\frac{1}{2}} s {\rm Im}\,(T_1 T_2)
\end{equation} 
($X_{12}$ $\!=$ $\!X_{21}$ $\!=$ $\!Y_{12}$ $\!=$ $\!Y_{21}$ $\!=$ $\!0$),
where $\sigma$ $\!=$ $\!+1(-1)$ for absorbing (amplifying) devices, 
and $n_{{\rm th}\,i}$ $\!=$ $\![\exp \hbar\omega/(k_B\vartheta_i)-1]^{-1}$ 
is the mean number of the (thermal) excitations of the $i$th device. 

Application of the Peres--Horodecki separability criterion
\cite{Simon00}
\begin{equation}
\label{3.8}
\det {\bf X} \det {\bf Y} + \left( \textstyle\frac{1}{4}
-|\det{\bf Z}| \right)^2 -{\rm Tr}\left( {\bf XJZJYJZ}^T{\bf J}
\right) \ge \textstyle\frac{1}{4} \left( \det{\bf X} +\det{\bf Y}
\right)
\end{equation}
with
\begin{equation}
{\bf J} = \left( \begin{array}{cc} 0&1\\-1&0 \end{array} \right)
\end{equation}
to the output variance matrix yields (for equal devices) the 
inequality 
\begin{equation}
\label{3.9}
n_{\rm th} \ge
\frac{\left( 1-\sigma \right) \left( 1-|R|^2 \right)
+|T|^2\left(\sigma-{\rm e}^{-2|\zeta|}\right)}{2\sigma \left(
1-|R|^2-|T|^2 \right)} \,.
\end{equation}
Hence, for chosen squeezing 
parameter $\zeta$ and chosen transmission and reflection coefficients 
$T$ and $R$, respectively, there exists a maximal temperature
and correspondingly a maximal mean excitation number of the thermal
state in which each of the two devices is prepared 
such that the quantum state of the transmitted squeezed vacuum is 
still not separable.

In particular, for absorbing devices ($\sigma$ $\!=$ $\!+1$) 
the inequality (\ref{3.9}) reads
\begin{equation}
\label{3.10}
n_{\rm th} \ge \frac{|T|^2 \left( 1-{\rm e}^{-2|\zeta|}\right)}
{2\left( 1-|R|^2-|T|^2 \right)}\,.
\end{equation}
With regard to optical fibers with perfect input coupling, we may let
$R$ $\!=$ $\!0$ and relate, according to the Lambert-Beer law,
the transmission coefficient to the propagation length $l$ as 
$|T|$ $\!=$ $\!e^{-l/l_{\rm A}}$, with $l_{\rm A}$ being the
characteristic absorption length of the fibers. From the
inequality (\ref{3.9}) it then follows that the upper bound of
the propagation length for which the transmitted squeezed vacuum
is still not separable is
\begin{equation}
\label{3.10a}
l = {\textstyle\frac{1}{2}}l_{\rm A}
\ln\!\left[1 + \frac{1}{2n_{\rm th}}\left(1-{\rm e}^{-2|\zeta|}
\right)\right]. 
\end{equation}
It is interesting to note that this result
agrees with that calculated in \cite{Lee00}, if the `renormalized
time' in \cite{Lee00} is replaced with $1$ $\!-$ $\!|T|^2$. 
Note that in our formalism the bound is
simply obtained  by using the quantum-optical input-output 
relations (\ref{1.1}) for light at dispersing and absorbing dielectric 
four-port devices. Involved calculations of the dynamics of the Wigner 
function are not needed. 

Moreover, the general result (\ref{3.9}) also applies to 
amplifying devices ($\sigma$ $\!=$ $\!-1$). In particular, in the
zero-temperature limit ($n_{\rm th}$ $\!=$ $\!0$)   
the boundary between inseparability and separability is reached when
\begin{equation}
\label{3.11}
|T|^2 = \frac{2\left( 1-|R|^2 \right)}{1+{\rm e}^{-2|\zeta|}}
\end{equation}
is valid. For zero reflection ($R$ $\!=$ $\!0$), Eq.~(\ref{3.11})
reveals that the upper limit of the 'excess' gain $g$ $\!=$
$\!|T|^2$ $\!-$ $\!1$ $\!\ge$ $\!0$ for which inseparability changes
to separability is simply given by the squeezing parameter $q$,
\begin{equation}
\label{3.12}
g = |q| = \tanh |\zeta| \,.
\end{equation}
An obvious consequence of Eq.~(\ref{3.12}) is that entanglement cannot 
be produced from the vacuum by amplification. For the vacuum the 
squeezing parameter has to be set equal to zero and thus from
Eq.~(\ref{3.12}) it follows that any nonvanishing gain $g$ must
necessarily lead to a separable state.
Another interesting fact is that there exists an absolute upper bound
of the gain for which inseparability can be retained. Since the absolute
value of the squeezing parameter (and thus the 'excess' gain $g$) is 
bounded by $1$, one is left with the conclusion that an amplifier 
which doubles the intensity of a signal ($|T|^2$ $\!=$ $\!2$) destroys 
all but the maximal entanglement in a two-mode squeezed vacuum 
state which spoils the use of fiber amplifiers in quantum communication.

\section{Entanglement estimates}
\label{estimate}

The separability criterion exploited in Sec.~\ref{sep}
can tell us only if the transmitted quantum state is separable 
or not. It can not, however, provide us information about the 
amount of entanglement which is actually contained in the state. 
In order to obtain analytical estimates of the available
amount of entanglement, we note that any entanglement measure 
(such as the distance of the state under consideration to the set of 
separable states measured by the relative entropy) has the 
convexity property \cite{Horodecki00}
\begin{equation}
\label{4.1}
E[\lambda\hat{\varrho}_1 \!+\!(1\!-\!\lambda)\hat{\varrho}_2] 
\le \lambda E(\hat{\varrho}_1) +(1-\lambda) E(\hat{\varrho}_2).
\end{equation}
This property can advantageously be used to find bounds on the
entanglement. If we are able to divide the quantum
state into a sum of separable states (having no entanglement) and a
single pure state, then an upper bound on the entanglement is given 
by the reduced von Neumann entropy of the extracted pure state
\cite{Scheel00}. 

Let us consider two modes that are initially prepared in 
a truncated version of the quantum state (\ref{0.1})
\begin{equation}
\label{4.2}
|\phi(q)\rangle = \frac{1}{\sqrt{1+q^2}} \left( |00\rangle +q|11\rangle 
\right) ,
\end{equation}
which approximates a two-mode squeezed vacuum for small values 
of the squeezing parameter (i.e., $|q|$ $\!\ll$ $\!1$).
It is not difficult to prove that the entanglement of the state is
\begin{equation}
\label{4.3}
E= \ln\! \left( 1+q^2 \right) - \frac{q^2}{1+q^2}\, \ln q^2 .
\end{equation}
Applying the transformation formula (\ref{1.8}), the quantum
state in which the two modes are prepared after propagating
through two fibers of transmission coefficients $T_1$ and
$T_2$ is derived to be
\begin{eqnarray}
\label{4.4}
\lefteqn{
\hat{\varrho}_{\rm out}^{({\rm F})} = \frac{q^2}{1+q^2} 
\left[
\left( 1-|T_1|^2 \right) \left( 1-|T_2|^2 \right)
|00\rangle\langle 00|
+ |T_1|^2 \left( 1-|T_2|^2 \right)
|10\rangle\langle 10|
\right. 
}
\nonumber \\ &&\hspace{10ex}
\left.
+\,|T_2|^2 \left( 1-|T_1|^2 \right)
|01\rangle\langle 01| 
\right]
+ \frac{1+|q'|^2}{1+|q|^2}\,|\phi(q')\rangle\langle\phi(q')|, 
\end{eqnarray}
where
\begin{equation}
\label{4.4a}
q' = T_1 T_2 q.
\end{equation}
Here and the following we assume that the fibers are prepared in the
ground state (low-temperature limit).
Since the first term on the right-hand side in Eq.~(\ref{4.4}) 
is a sum of separable
states and the second term is a pure state whose entanglement
is given by Eq.~(\ref{4.3}) with $q'$ in place of $q$,
the entanglement of the state in Eq.~(\ref{4.4}) can be estimated,
on recalling the convexity property (\ref{4.1}), according to 
\begin{equation}
\label{4.5}
E \le \frac{1}{1+q^2} \left[ \left( 1+|qT_1T_2|^2 \right) \ln \left(
1+|qT_1T_2|^2 \right) -|qT_1T_2|^2 \ln |qT_1T_2|^2 \right] .
\end{equation}
We see that with increasing propagation lengths, i.e., with
decreasing transmission coefficients, the entanglement of the
transmitted light decreases more rapidly than the intensity. 

Now let us return to the exact two-mode squeezed vacuum. 
Applying Eq.~(\ref{1.8}) and
transforming the density operator $\hat{\varrho}_{\rm in}$
$\!=$ $\!|\psi\rangle\langle\psi|$ with $|\psi\rangle$ from
Eq.~(\ref{0.1}), we derive after a lengthy, but straightforward 
calculation
\begin{eqnarray}
\label{4.7}
\lefteqn{
\hspace{-5ex}
\hat{\varrho}_{\rm out}^{({\rm F})} = \left( 1-q^2 \right)
\sum\limits_{m=0}^\infty \sum\limits_{n=0}^m \frac{1}{m!n!}
\left[ \sum\limits_{k=0}^\infty C_{m,n,k} \left( c_{m-n}
|m-n+k\rangle \langle k| +\mbox{H.c.} \right) \right] 
}
\nonumber \\[.5ex] && \hspace{10ex}
\otimes
\left[ \sum\limits_{l=0}^\infty D_{m,n,l} \left( d_{m-n}
|m-n+l\rangle \langle l|  
+\mbox{H.c.} \right)\right] ,
\end{eqnarray}
where
\begin{eqnarray}
\label{4.8a}
C_{m,n,k} &=& \frac{q^n m!n!\left( 1-|T_1|^2 \right)^{n-k}
|T_1|^{2k}}{(n-k!)\sqrt{k!(m-n-k!)}} \,,
\\[.5ex]
\label{4.8b}
c_{m-n} &=& q^{(m-n)/2} T_1^{m-n} 
\left( 1-{\textstyle\frac{1}{2}}\delta_{mn}
\right) 
\\[.5ex]
\label{4.8c}
D_{m,n,l} &=& \frac{q^n m!n!\left( 1-|T_2|^2 \right)^{n-l}
|T_2|^{2l}}{(n-l!)\sqrt{l!(m-n-l!)}} \,,
\\[.5ex]
\label{4.8d}
d_{m-n} &=& q^{(m-n)/2} T_2^{m-n} 
\left( 1-{\textstyle\frac{1}{2}}\delta_{mn}
\right) \,.
\end{eqnarray}
Performing the sum over $n$ then yields the final result
\begin{equation}
\label{4.9}
\hat{\varrho}_{\rm out}^{({\rm F})} = \left( 1-q^2 \right)
\sum\limits_{m=0}^\infty \sum\limits_{k,l=0}^\infty K_{k,l,m}
\left( c_m |m+k\rangle\langle k| +\mbox{H.c.} \right) \otimes
\left( d_m |m+l\rangle\langle l| +\mbox{H.c.} \right),
\end{equation}
where
\begin{eqnarray}
\label{4.10}
\lefteqn{
K_{k,l,m} = \frac{[q^2 (1\!-\!|T_1|^2) (1\!-\!|T_2|^2)]^a
a!(a\!+\!m)!}{\sqrt{k! l! (k\!+\!m)!(l\!+\!m)!}(a\!-\!k)!(a\!-\!l)!} 
\left( \frac{|T_1|^2}{1\!-\!|T_1|^2} \right)^{k}
\left( \frac{|T_2|^2}{1\!-\!|T_2|^2} \right)^{l}  
}
\nonumber\\[.5ex] && \hspace{8ex}
\times\, {}_3F_2\! \left[ \begin{array}{c}
a+1, a+m+1,1\\ a-k+1, a-l+1 \end{array} \!;
q^2 \left(1-|T_1|^2\right)\left(1-|T_2|^2\right) \right] 
\end{eqnarray}
[$a$ $\!=$ $\!\max(k,l)$]. 
Note that one lower index of the
hypergeometric function ${}_3F_2$ is always equal to unity, so that
we effectively deal with a Gaussian hypergeometric function ${}_2F_1$.

We now try to decompose the density operator (\ref{4.9}) 
into a pure state $|\Psi\rangle$ and some residual state  
$\hat{\varrho}'$ whose entanglement is desired to be sufficiently small,
\begin{equation}
\label{4.10a}
\hat{\varrho}_{\rm out}^{({\rm F})} 
= \lambda \hat{\varrho}' + (1-\lambda)|\Psi\rangle\langle\Psi|,
\end{equation}
A suitable pure state $|\Psi\rangle$ may be chosen such that 
\begin{equation}
\label{4.11}
\sqrt{1-\lambda}\, |\Psi\rangle 
= \sqrt{\frac{1-q^2}{K_{000}}} \sum\limits_{n=0}^\infty
K_{00n} c_n d_n |nn\rangle .
\end{equation}
It has the properties that (i) only matrix elements 
of the same type as in the initial squeezed vacuum occur and (ii) 
the coefficients of the matrix elements \mbox{$|00\rangle$ 
$\!\leftrightarrow$ $\!|nn\rangle$} are met exactly, i.e.,
\begin{equation}
\label{3.11a}
(1-\lambda)\langle 00|\Psi\rangle\langle\Psi|nn\rangle
=\langle 00|\hat{\varrho}_{\rm out}^{\rm (F)}|nn\rangle. 
\end{equation} 
This choice is of course not unique. 
Moreover, the residual state might still contain some entanglement. 
For small enough squeezing parameter $q$, however, the 
residual entanglement is expected to be small compared to the 
entanglement contained in the state (\ref{4.11}), i.e.,
$\lambda E(\hat{\varrho}')$ $\!\ll$ $\!(1$ $\!-$
$\!\lambda)E(|\Psi\rangle)$. In principle, one can proceed and extract more 
and more pure 
states from the output state and apply the generalized inequality
\begin{equation}
\label{4.12}
E\Big( \sum\limits_i p_i\hat{\varrho}_i \Big) 
\le \sum\limits_i p_i \, E\!\left(\hat{\varrho}_i\right) ,
\qquad \sum\limits_i p_i =1 .
\end{equation}

Disregarding a possible (small) entanglement of the residual
state $\hat{\varrho}'$, the entanglement of the pure state $|\Psi\rangle$
gives us some estimate of (the upper bound of) entanglement of the 
output state $\hat{\varrho}_{\rm out}^{\rm (F)}$,   
\begin{eqnarray}
\label{4.13}
\lefteqn{
E(\hat{\varrho}_{\rm out}^{({\rm F})}) 
\approx (1-\lambda)\,E(|\Psi\rangle) 
}
\nonumber\\&&
=\frac{1\!-\!x}{(1\!-\!x)^2\!-\!y}\,
\ln\!\left[\frac{1\!-\!x}{(1\!-\!x)^2\!-\!y}\right] 
+ \frac{(1\!-\!x)\{[y\!+\!(1\!-\!x)^2] 
\ln (1\!-\!x)\!-\!y \ln y\}}{[y\!-\!(1\!\!-x)^2]^2} \,,
\end{eqnarray}
where
\begin{equation}
\label{4.13a}
x = q^2 (1-|T_1|^2) (1-|T_2|^2) ,
\end{equation}
\begin{equation}
\label{4.13b}
y = |qT_1T_2|^2 .
\end{equation}
Note that for $T_1$ $\!=$ $\!T_2$ $\!=$ $\!1$ Eq.~(\ref{4.13})
gives the correct entanglement of the input state,
\begin{equation}
\label{4.14}
\left.E(\hat{\varrho}_{\rm out}^{\rm (F)})\right|_{T_1=T_2=1}
= E(|\psi\rangle) 
= - \ln \left( 1-q^2 \right)
-\frac{q^2}{\left( 1-q^2 \right)} \ln q^2. 
\end{equation}

\begin{figure}[h]
\hspace*{3.5cm}
\psfig{file=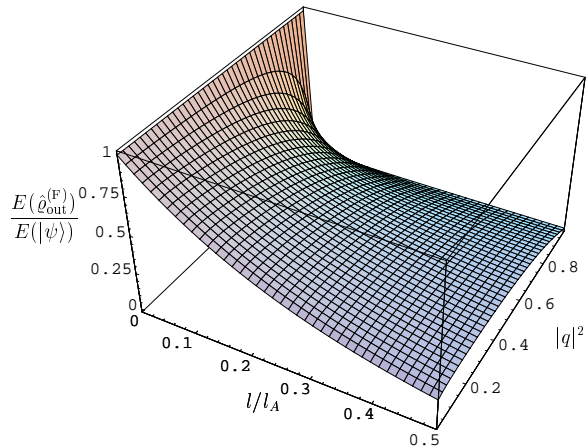,width=8cm}
\caption{\label{bild} 
\protect\parbox[t]{.67\textwidth}{
Estimate of the entanglement of the transmitted squeezed
vacuum, Eq.~(\protect\ref{4.13}),
as a function of the initial squeezing parameter $|q|^2$
and the transmission length $l$
[$E(|\psi\rangle)$, initial entanglement].}
}
\end{figure}

\begin{figure}[h]
\hspace*{3.5cm}
\psfig{file=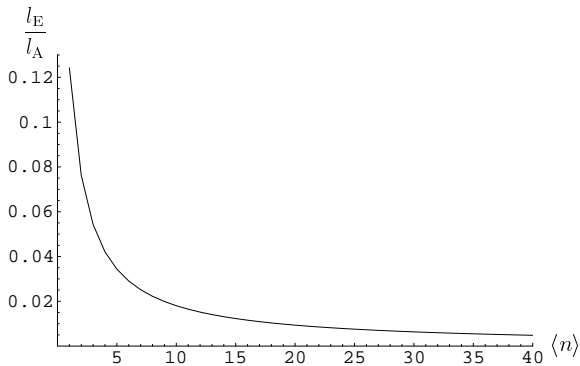,width=8cm}
\caption{\label{deg} 
\protect\parbox[t]{.67\textwidth}{
Entanglement degradation length \protect$l_{\rm E}$
after which the entanglement of the transmitted two-mode squeezed
vacuum has decreased to half the initial value as a function of 
the initial value of the mean number of photons $\langle n\rangle$
($l_{\rm A}$, absorption length).}
}
\end{figure}

In Fig.~\ref{bild} we have illustrated the dependence of the
estimated entanglement of the transmitted two-mode squeezed vacuum,
Eq.~(\ref{4.13}), on both the fiber length and the strength of 
the initial squeezing.   
It is seen that with increasing strength of the initial squeezing
[which is, according to Eq.~(\ref{4.14}), a measure of the 
strength of the initial entanglement] the entanglement of the 
transmitted light drastically decreases with the transmission 
length. The transmission length $l_{\rm E}$ at which the entanglement
degradation has reached half of the initial entanglement
is shown in Fig.~\ref{deg} as a function of the mean number
of initial photons $\langle n\rangle$, which is related to
the squeezing parameter $|q|^2$ according to
\begin{equation}
\label{4.15}
q^2 = \frac{\langle n\rangle}{\langle n\rangle +1} \,.
\end{equation}
Figure \ref{deg} reveals that even for relatively small mean
number of photons this characteristic length of entanglement 
degradation is much more shorter than the absorption length. 
Obviously, entanglement cannot be
maintained when going to more macroscopic nonclassical states.
Since strong squeezing (i.e., large photon number)
is typically required in quantum teleportation \cite{Braunstein98},
one has to find a compromise between highest
possible entanglement and lowest entanglement 
degradation.

\section{Summary and Outlook}
\label{summary}

We have studied the entanglement degradation
of a two-mode squeezed vacuum state transmitted through
noisy communication channels, a typical example being the
propagation along dispersing and absorbing or amplifying
optical fibers. As expected, both absorption and
amplification lead to entanglement degradation, because of
the additional noise introduced by them.
Using the quantum-optical input-output relations of radiation 
at dielectric four-port devices, we have derived the maximal
transmission length after which an initially entangled state becomes
separable. Analogously, we have found that there is 
a maximal gain factor for which inseparability (of any two-mode
squeezed vacuum state) can be retained at all. 

Knowledge of the complete quantum state at the output
of the device enables us, in principle, to compute 
the amount of entanglement available after transmission
such as the distance of the state to the set of all
separable states. 
However, for higher-dimensional Hilbert spaces as
in our case this is impossible to do in general. Therefore we have  
restricted ourselves to the calculation of upper bounds or 
estimates of upper bounds. For
weak squeezing we may truncate the Hilbert space effectively at low
photon numbers providing us with a way to establish an upper
bound on the entanglement by exploiting the convexity property
of entanglement measures. The procedure is based on the extraction of
a single pure state from the output state  leaving behind only
separable states. For the general case of infinite dimensional
Hilbert space we have derived an estimate of the upper
bound.  

Still, we are left with (estimates) of bounds on the
entanglement. Future works must surely contain algorithms for
computation of the entanglement, at least for Gaussian states. 
A possible step in that direction would include the
calculation of the distance of a given Gaussian state to the set of
all separable Gaussian states whose surface can again be parametrized
by the Peres--Horodecki separability criterion.


\end{document}